\documentstyle[12pt,epsfig]{article}

\def\e{\epsilon}

\def\e3{$\epsilon_3$}

\def\ch2{$\chi^2$}

\def\co#1{{\ifmmode{\cal O}_{#1}\else${\cal O}_{#1}$\fi}}

\newdimen\unit
\def\point#1 #2 #3{\vbox to0pt{\kern-#2\unit
  \hbox{\kern#1\unit#3}\vss}
 \nointerlineskip}

\newcommand{\be}{\begin{equation}}
\newcommand{\ee}{\end{equation}}
\newcommand{\bea}{\begin{eqnarray}}
\newcommand{\eea}{\end{eqnarray}}

\begin{document}
\thispagestyle{empty} \noindent
\begin{flushright}
        OHSTPY-HEP-T-03-001  \\
                February 2003
\end{flushright}

\vspace{1cm}
\begin{center}
  \begin{Large}
  \begin{bf}
A Natural Framework for Bi-large Neutrino Mixing

\end{bf}
 \end{Large}
\end{center}
  \vspace{1cm}
    \begin{center}
 S. Raby\\
      \vspace{0.3cm}
\begin{it}
Department of Physics, The Ohio State University, \\ 174 W. 18th
Ave., Columbus, Ohio  43210
\end{it}
  \end{center}
  \vspace{1cm}
\centerline{\bf Abstract}
\begin{quotation}
\noindent In this letter we present a ``natural" framework for
obtaining bi-large neutrino mixing incorporating the
Frampton-Glashow-Yanagida neutrino mass matrix ansatz. We show
that an $SU(2) \times U(1)$ family symmetry can provide the
desired FGY neutrino mass ansatz in a minimal supersymmetric
standard model. We also show how to obtain an approximate FGY
ansatz in an $SO(10)$ SUSY GUT. In this context, the same $SU(2)
\times U(1)$ family symmetry also generates the hierarchy of
fermion masses as well as ameliorating SUSY flavor problems.
\end{quotation}

\newpage

In a recent paper \cite{Frampton:2002qc} [FGY] a simple ansatz for
neutrino masses has been presented which has several significant
virtues. It is an elegant framework for so-called bi-large
neutrino mixing, i.e. giving maximal $\nu_\mu - \nu_\tau$ mixing
for atmospheric neutrinos and large mixing angle
Mikheyev-Smirnov-Wolfenstein mixing for solar
neutrinos.\footnote{For other recent discussions on {\em similar}
frameworks for bi-large mixing, see Refs.
\cite{Kuchimanchi:2002yu,King:2002qh}}  In addition, the sign of
the cosmological baryon asymmetry is related to CP violation which
may be observable in low energy neutrino oscillation
experiments.\footnote{For some recent discussions of leptogenesis
and low energy physics, see Refs.
\cite{Ellis:2002xg}-\cite{Branco:2002xf}.} In this letter we
present a simple $SU(2) \times U(1)$ symmetry within the context
of supersymmetry which makes the FGY ansatz  ``natural". We then
extend this simple realization to an $SO(10)$ grand unified theory
[GUT] which also fits charged fermion masses and mixing angles. In
the latter example,  the possibility of additional CP violating
angles destroys the leptogenesis/neutrino CP violation connection,
but still provides a natural framework for bi-large neutrino
mixing.

Consider the three lepton doublets  $l_i,  \ i = 1,2,3$. Under an
$SU(2)$ family symmetry, two of the lepton doublets transform as a
doublet given by $L_a = \left(\begin{array}{c} l_1 \\
l_2 \end{array} \right)$, while $l_3$ is a singlet.  In addition,
we require two flavor (anti)-doublets  $\phi^a$ and $\tilde
\phi^a$, 4 flavor singlets $N_1, \ N_2, \ S_1, \ S_2$ and the
standard model Higgs doublet $H$.

The superpotential is given by
\begin{equation} W = \frac{H}{M} \ ( \ L_a \ \phi^a \ N_1 \ +
 \ L_a \ \tilde \phi^a \ N_2 \  +  \ l_3 \ \omega \ N_2 \ ) \ + \
\frac{1}{2} \ ( \ S_1 \ N_1^2 \ + \ \ S_2 \ N_2^2 \ )
\end{equation} where $M$ is some large mass scale.  Note
arbitrary order one dimensionless couplings are implicitly assumed
to multiply the terms in $W$.   In addition, we assume that $W$ is
CP invariant where CP is presumed to be spontaneously violated in
complex vevs.  The $U(1)$ charge assignments for the fields $\{
L_a, \ l_3, \ $ $ N_1, \ N_2,\ $ $ \phi^a, \ \tilde \phi^a, \ $
$\omega, \ S_1, \ S_2 \}$ are as follows $\{ 1,\ \alpha,\ $ $x,\
y,\ $ $ -(x+1),\ -(y + 1),\ $ $-(\alpha + y), \ -2 x,\ -2 y \}$
with $x \neq y$.

We assume $\phi, \ \tilde \phi$ get vevs $\langle \phi \rangle =
\left(\begin{array}{c} \phi^1 \\ \phi^2
\end{array}
\right)$ and $\langle \tilde \phi \rangle = \left(\begin{array}{c} 0 \\
\tilde \phi^2 \end{array} \right)$.  This does not require fine
tuning since any arbitrary vevs can be rotated into this direction
using an $SU(2) \times U(1)$ flavor rotation. The vevs $\langle
S_i \rangle = M_i, \; i = 1,2$ are also needed to give the states
$N_i$ a large see-saw mass.   These vevs can be obtained via
suitable terms added to the superpotential.  We then obtain the 3
x 2 mixing matrix $D^{tr}$ defined by
\begin{equation}  \left(\begin{array}{ccc} \nu_1 & \nu_2 & \nu_3 \end{array} \right) \ D^{tr} \
\left( \begin{array}{c}  N_1 \\ N_2 \end{array} \right) \; \equiv
\;  \left(\begin{array}{ccc} \nu_1 & \nu_2 &
\nu_3 \end{array} \right) \left(\begin{array}{cc} a & 0 \\
a^\prime \ e^{i \delta} & b \\ 0 & b^\prime \end{array} \right)
\left(
\begin{array}{c}  N_1 \\ N_2 \end{array} \right) \end{equation}
where \begin{equation}  a \;  = \;  v \sin\beta \
\frac{\phi^1}{\sqrt{2} M}, \;\;\;  a^\prime \ e^{i \delta} \; = \;
v \sin\beta \ \frac{\phi^2}{\sqrt{2} M},  \;\;\;   b \; = \; v
\sin\beta \ \frac{\tilde \phi^2}{\sqrt{2} M}, \;\;\; b^\prime \; =
\; v \sin\beta \ \frac{\omega}{\sqrt{2} M} \label{eq:phase}
\end{equation} and $ \langle H \rangle \; = \;
\left(\begin{array}{c} 0 \\ v \sin\beta/\sqrt{2}
\end{array} \right)$.    Note, the vevs are in general complex,
however, with phase redefinitions of the fields $N_1, \ N_2, \
l_1, \ l_2, \ l_3$  all phases but $\delta$ can be transformed
away.  As we show later, this is consistent with the charged
lepton sector.

The neutrino masses are given by
\begin{equation} {\cal M}_{FGY} = D^{tr} \ M_N^{-1} \ D \end{equation}
with
\begin{equation} M_N = \left(
\begin{array}{cc} M_1 & 0
\\ 0 & M_2 \end{array} \right) . \end{equation}   This is exactly
the FGY ansatz \cite{Frampton:2002qc}.   The diagonalized neutrino
mass matrix is given by
\begin{equation} {\cal M}^D = U_{FGY}^{tr} \ {\cal M}_{FGY} \ U_{FGY} \end{equation}
where \begin{equation} U_{FGY} = \left( \begin{array}{ccc}
1/\sqrt{2} &  1/\sqrt{2} & 0 \\ -1/2 & 1/2 & 1/\sqrt{2} \\
1/2 & -1/2 & 1/\sqrt{2} \end{array} \right) \times \left(
\begin{array}{ccc}
1 &  0 & 0 \\ 0 & cos\theta & sin\theta \\
0 & -sin\theta & cos\theta \end{array} \right) \label{eq:ufgy}
\end{equation} is the FGY neutrino mixing matrix (for the case $b
= b^\prime$ and $a^\prime = \sqrt{2} \ a$). The three neutrino
mass eigenvalues are given by
\begin{equation} m_{\nu_3} \approx  2 b^2/M_2 \hspace{.2in} \gg \hspace{.2in}
m_{\nu_2} \approx 2 a^2/M_1, \hspace{.2in} m_{\nu_1} = 0
\label{eq:numass}
\end{equation} and the small angle $\theta \sim
m_{\nu_2}/(\sqrt{2} \ m_{\nu_3})$.  Recall, in order to fit
neutrino data we have
\begin{eqnarray}  m_{\nu_3} \approx & 0.05 \;\; {\rm eV} =
& \sqrt{\Delta m^2_{atm}} \\
m_{\nu_2} \approx & 7 \times 10^{-3} \;\; {\rm eV} = &
\sqrt{\Delta m^2_{sol}} \nonumber \end{eqnarray}

In order to complete the discussion of neutrino masses and mixing
angles, it is necessary to discuss the charged lepton mass matrix.
If this matrix is to a good approximation diagonal, then the
neutrino mass matrix above may be considered to be in a lepton
flavor basis and our results are unaffected by the charged lepton
sector.   There are many possible choices for the $SU(2) \times
U(1)$ assignments of the right handed charged leptons.   In order
to be more specific we shall discuss an example in which the
charged leptons are $SU(2)$ singlets.   Consider the
superpotential
\begin{equation} W_{ch. \; leptons} = \frac{\bar H}{M} \ ( \ L_a \ \phi^a \ \bar e_1 \ +
 \ L_a \ \tilde \phi^a \ \bar e_2 \ +  \ l_3 \ ( \ \omega \ \bar e_2 \ + \ \bar \omega \ \bar
 e_3 \ )) \label{eq:chleptons} \end{equation} where \begin{equation}  \bar a \;  = \;  v \cos\beta \
\frac{\phi^1}{\sqrt{2} M}, \;\;\;  \bar a^\prime \ e^{i \delta} \;
= \; v \cos\beta \ \frac{\phi^2}{\sqrt{2} M},  \;\;\;   \bar b \;
= \; v \cos\beta \ \frac{\tilde \phi^2}{\sqrt{2} M},
\end{equation}  \begin{equation} \bar b^\prime \;
= \; v \cos\beta \ \frac{\omega}{\sqrt{2} M}, \;\;\; \bar c = v
\cos\beta \ \frac{\bar \omega}{\sqrt{2} M} .
\end{equation} and $ \langle \bar H \rangle \; = \; \left(
\begin{array}{c}  v \cos\beta/\sqrt{2} \\ 0 \end{array} \right). $
The $U(1)$ charge assignments for the fields $\{ \bar e_1, \ \bar
e_2, \ $ $ \bar e_3, \ \bar \omega \}$ are as follows $\{ x ,\ y,\
$ $z, \ -(\alpha + z) \}$.  Note, with suitable phase
redefinitions of the fields $\bar e_i, \; i = 1,2,3$, all phases
but $\delta$ may be transformed away.

The charged lepton mass matrix is then given by (defined with
left-handed doublets on the right)
 \begin{equation} m_l = \left( \begin{array}{ccc}
\bar a & \bar a^\prime \ e^{i \delta} & 0 \\ 0 & \bar b & \bar b^\prime \\
0 & 0 & \bar c \end{array} \right)  . \label{eq:mchleptons}
\end{equation}   Note,  using the constraint from the neutrino
sector, we have $\bar a \approx \bar a^\prime$,  $\bar b \approx
\bar b^\prime$ and we take $\bar a, \; \bar a^\prime \ll \bar b,
\; \bar b^\prime \ll  \bar c$. Hence   \begin{equation} m_e
\approx \bar a, \;\;\; m_\mu \approx \bar b, \;\;\; m_\tau \approx
\bar c \label{eq:chlepmass}
\end{equation} and the charged lepton mass matrix is approximately
diagonal. The matrix can be diagonalized by a left-handed
(right-handed) unitary transformation $U_e = {\cal P} \ V_e$
($U_{\bar e} = {\cal P} \ V_{\bar e}$) with $m_l^{diagonal} =
U_{\bar e}^\dagger  \ m_l \ U_e$ where  $V_e$ ($V_{\bar e}$) are
real orthogonal matrices and ${\cal P}$ is a diagonal phase
matrix. We have
\begin{equation} U_e \approx   \left(
\begin{array}{ccc} 1 &  0 & 0 \\ 0 & e^{-i \delta} & 0 \\
0 & 0 & -e^{-i \delta} \end{array} \right)   \times  \left(
\begin{array}{ccc} 1 &  \bar a^2/\bar b^2 & 0 \\ -\bar a^2/\bar
b^2 & 1 & -\bar b^2/\bar c^2 \\ 0 & \bar b^2/\bar c^2 & 1
\end{array} \right) \approx V_e \ {\cal P} \label{eq:ve}
\end{equation} where $V_e$ is approximately the $3 \times
3$ unit matrix with negligible mixing angles. In the lepton flavor
basis we have
\begin{equation} {\cal M} =  U_e^{tr} \; [ \ D^{tr} \ M_N^{-1} \ D \ ] \;
U_e  \approx {\cal M}_{FGY}  \end{equation}   where with a
suitable phase redefinition of the lepton doublets and singlets we
recover the FGY ansatz.

It is clear that we have obtained the FGY ansatz as a consequence
of a spontaneously broken $SU(2) \times U(1)$ family symmetry.
Note that the FGY ansatz has a discrete ambiguity related to the
mass ordering of the heavy Majorana neutrinos.  It has been shown
\cite{Raidal:2002xf} that leptogenesis and the observed baryon
asymmetry correlates this ambiguity with the sign of the CP
violating phase $\delta$ such that for $M_1 \ll M_2 \;\; ( \delta
> 0)$ or for $M_1 \gg M_2 \;\; (\delta < 0)$. These two
possibilities also lead to different predictions for low energy
lepton flavor violation \cite{Raidal:2002xf}. This ambiguity is
resolved in our model since now (using Eqns. \ref{eq:numass} and
\ref{eq:chlepmass}) we have $ (m_e/m_\mu)^2 \approx (\bar a/\bar
b)^2 \approx (a/b)^2 \approx (M_1/M_2) (m_{\nu_2}/m_{\nu_3})$.
Hence $(M_1/M_2) \sim 10^{-3}$.  We thus have the predictions for
the lepton flavor violating branching ratios \cite{Raidal:2002xf}:
\begin{eqnarray} B( \mu \rightarrow e \gamma) \approx & 2 \ r \
10^{-13} & \label{eq:flavor} \\ B( \tau \rightarrow \mu \gamma)
\geq & 3 \ r \ 10^{-12} & \nonumber
\end{eqnarray} with $ r \approx (\tan\beta/10)^2 \ (150 \ {\rm GeV}/
m_{SUSY})^4$.

Note, the FGY ansatz can accomodate bi-maximal mixing but it
cannot explain why $a \sim a^\prime$ or $b \sim b^\prime$. This
difficulty, however, is somewhat reduced in our $SO(10)$ example.
In addition, note this model has a straightforward generalization
to $SU(5)$ with the conversions of $L_a, \ l_3 \rightarrow \bar
5_a, \ \bar 5_3$; $\; \bar e_i \rightarrow 10_i, \; i = 1,2,3$ and
the Higgs doublets $H \rightarrow 5_H, \ \bar H \rightarrow \bar
5_H$. Of course we would then need to discuss quark masses and
resolve the problem of the simple GUT scale $SU(5)$ relations --
$\lambda_d = \lambda_e, \ \lambda_s = \lambda_\mu$.   However, we
will not consider an $SU(5)$ extension further here. Instead we
consider a simple extension to $SO(10)$. However before we discuss
an $SO(10)$ model, let us consider the family symmetry $SU(2)
\times U(1)$ further and possible family symmetry breaking
mechanisms. Perhaps the first question to address is whether the
family symmetry is local or global.  If, however, we assume that
the scale of symmetry breaking $M$ is sufficiently large with $M
>> M_Z$ , then either way it will not have any observable low
energy consequences. If it is a global symmetry, the massless
Nambu-Goldstone bosons will decouple from low energy physics as
$1/M$, whereas if it is a local symmetry the new $SU(2) \times
U(1)$ gauge bosons will get mass of order $M$, hence they also
decouple from the low energy physics. On the other hand, the
question of whether it is a local or global symmetry will affect
the symmetry breaking mechanisms. Consider for example that it is
a local symmetry. Then in order to find D flat directions, we may
need to include conjugate $SU(2)$ doublet flavon fields $\phi^c,
\; \tilde \phi^c$ with $U(1)$ charges $(x + 1), \;\; (y + 1)$.
These then get non-zero vevs in the D flat directions.   In
addition, if the $U(1)$ has a non-vanishing trace, it will obtain
a Fayet-Iliopoulos D term of order the effective cut-off scale of
the theory.  This might give these flavon fields a supersymmetry
preserving vev at this scale.

Now consider a similar realization of the same symmetry in an
$SO(10)$ SUSY GUT with an $SU(2) \times U(1)$ family
symmetry.\footnote{For related charged fermion analyses in
$SO(10)$ SUSY GUTS with $SU(2) \times U(1)^n$ (or $D_3$) family
symmetries, see \cite{Barbieri:1996ww,Blazek:1999ue} (or
\cite{Dermisek:1999vy}).}  The three families of quarks and
leptons are contained in three 16 dimensional representations of
$SO(10)$ $ \{ 16_a, \; 16_3 \}$ with $16_a, \; a = 1,2$ an $SU(2)$
flavor doublet. Consider the charged fermion sector first.
Although the charged fermion sector is not the main focus of this
letter it is necessary to present a possible superpotential which
results in charged fermion masses and mixing angles, since in
$SO(10)$ the neutrino and charged fermion sectors are inextricably
intertwined.

The superpotential resulting in charged fermion masses and mixing
angles is given by
\begin{eqnarray} W_{ch. fermions} = & 16_3 \ 10 \ 16_3 +  16_a \ 10 \
\chi^a & \label{eq:chfermion} \\ & +  \bar \chi_a \ ( M_{\chi} \
\chi^a + \ 45 \ \frac{\phi^a}{\hat M} \  16_3 \ + \ 45 \
\frac{\tilde \phi^a \ \tilde \phi^b}{(\hat M)^2} \  16_b + A^{a b}
\ 16_b ) & \nonumber
\end{eqnarray} where $M_{\chi} = M_0 ( 1 + \alpha X + \beta Y )$
includes $SO(10)$ breaking vevs in the $X$ and $Y$ directions,
$\phi^a, \ \tilde \phi^a, \ A^{a b} = -A^{b a}$ are $SO(10)$
singlet flavon fields with explicit $SU(2)$ properties, $\hat M$
is an $SO(10)$ singlet mass and $\langle 45 \rangle \sim (B - L) \
M_G$. The fields $\phi, \ \tilde \phi$ are assumed to obtain vevs
\begin{equation} \langle \phi \rangle =  \left( \begin{array}{c} \phi^1 \\
\phi^2 \end{array} \right), \; \langle \tilde \phi \rangle =
\left( \begin{array}{c} 0 \\ \tilde \phi^2 \end{array}
\right).\end{equation}  Note, once again, this is completely
general, since arbitrary vevs can be rotated into this direction
by the $SU(2) \times U(1)$ family symmetry.

The superpotential, (Eqn. \ref{eq:chfermion}) results in the
following charged fermion Yukawa matrices:\footnote{Note, in this
case the doublets are on the left.}
\begin{eqnarray}
Y_u =&  \left(\begin{array}{ccc}  0 & \epsilon' \ \rho & - \epsilon \ \xi  \\
             - \epsilon' \ \rho &  \tilde \epsilon \ \rho & - \epsilon     \\
       \epsilon \ \xi   & \epsilon & 1 \end{array} \right) \; \lambda & \nonumber \\
Y_d =&  \left(\begin{array}{ccc}  0 & \epsilon'  & - \epsilon \ \xi \ \sigma \\
- \epsilon'   &  \tilde \epsilon  & - \epsilon \ \sigma \\
\epsilon \ \xi  & \epsilon & 1 \end{array} \right) \; \lambda & \label{eq:yukawa} \\
Y_e =&  \left(\begin{array}{ccc}  0 & - \epsilon'  & 3 \ \epsilon \ \xi \\
          \epsilon'  &  3 \ \tilde \epsilon  & 3 \ \epsilon  \\
 - 3 \ \epsilon \ \xi \ \sigma  & - 3 \ \epsilon \ \sigma & 1 \end{array} \right) \; \lambda &
 \nonumber \end{eqnarray}
with  \begin{eqnarray}  \xi \;\; =  \;\; \phi^1/\phi^2; & \;\;
\tilde \epsilon  \;\; \propto   \;\; (\tilde \phi^2/\hat M)^2;  & \label{eq:omega} \\
\epsilon \;\; \propto  \;\; \phi^2/\hat M; &  \;\;
\epsilon^\prime \;\; \sim  \;\;  (A^{1 2}/M_0); \nonumber \\
  \sigma \;\; =   \;\; \frac{1+\alpha}{1-3\alpha}; &  \;\; \rho \;\; \sim   \;\;
  \beta \ll \alpha .& \nonumber \end{eqnarray}
It has been shown in Ref. \cite{Blazek:1999ue} that excellent fits
to charged fermion masses and mixing angles are obtained with this
Yukawa structure.

In the three $16$s we have three electroweak doublet neutrinos
($\nu_a, \ \nu_3$) and three electroweak singlet anti-neutrinos
($\bar \nu_a, \ \bar \nu_3$).\footnote{In an equivalent notation,
we have three left-handed neutrinos ($\nu_{L a} \equiv \nu_a, \;
\nu_{L 3} \equiv \nu_3$) and three right-handed neutrinos defined
by ($\nu_{R a} \equiv \bar \nu_a^*, \ \nu_{R 3} \equiv \bar
\nu_3^*$).}  The superpotential $W_{ch. fermions}$ also results in
a neutrino Yukawa matrix:
\begin{eqnarray}
Y_{\nu} =&  \left(\begin{array}{ccc}  0 & - \epsilon' \ \omega & {3 \over 2} \ \epsilon \ \xi \ \omega \\
      \epsilon'  \ \omega &  3 \ \tilde \epsilon \  \omega & {3 \over 2} \ \epsilon \ \omega \\
       - 3 \ \epsilon \ \xi \ \sigma   & - 3 \ \epsilon \ \sigma & 1 \end{array} \right) \; \lambda &
 \end{eqnarray} with $\omega \;\; =  \;\; 2 \, \sigma/( 2 \,
\sigma - 1)$ and a Dirac neutrino mass matrix given by
 \begin{equation} m_\nu \equiv Y_\nu \frac{v}{\sqrt{2}} \sin\beta.
 \label{eq:mnu}
  \end{equation}
 In addition,  the anti-neutrinos get GUT scale masses by mixing with three
$SO(10) \times SU(2)$ singlets $\{ N_i, \ i = 1,2,3 \}$. The full
superpotential is given by $W = W_{ch. fermions} + W_{neutrino}$
with
\begin{eqnarray} W_{neutrino} = & \frac{\overline{16}}{\hat M}  \left( N_1 \
\tilde \phi^a \ 16_a \ + \ N_2 \ \phi^a \ 16_a \ + \ N_3 \ \theta
\ 16_3 \right) & \\ & + \frac{1}{2} \left( S_1 \ N_1^2 \ + \ S_2 \
N_2^2 \right)  & \nonumber
\end{eqnarray}
and we assume $\overline{16}$ obtains a vev $v_{16}$ in the
right-handed neutrino direction, $\langle S_i \rangle = M_i$ for
$i = 1,2$ and $ \langle \theta \rangle = \theta$. We thus obtain
the effective neutrino mass terms given by
\begin{equation} W =  \nu \ m_\nu \ \bar \nu + \bar \nu \ V \ N +
\frac{1}{2} \ N \ M_N \ N \end{equation} with
\begin{equation} V^{tr} = \frac{v_{16}}{\hat M} \ \left(
\begin{array}{ccc} 0 &  \tilde \phi^2 & 0 \\
 \phi^1 & \phi^2 & 0 \\ 0 & 0 &  \theta \end{array}
\right), \; M_N = diag( M_1, M_2 , 0) \end{equation} where
$V^{tr}$ is the transpose of $V$.  The family symmetry is at least
$SU(2) \times U(1)$ where the $SU(2)$ charges are evident, while
the $U(1)$ charge assignments for $\{ \overline{16}, \ 16_3, \
16_a, \ $ $N_1, \ N_2, \ N_3,\ $ $\phi^a, \ \tilde \phi^a,\ $
$\theta, \ S_1, \ S_2 \}$ are given by $\{ -2 \ n_1 + n_2 - 1,\
1,\ x,\ $ $n_1,\ n_2,\ n_3,\ $ $2 \ (n_1 - n_2) + 1 - x, \ (n_1 -
n_2) + 1 - x,\ $ $ 2 \ n_1 - (n_2 + n_3), \ -2 n_1, \ -2 n_2 \}$.

The electroweak singlet neutrinos $ \{ \bar \nu, N \} $ have large
masses of order $V , M_N \sim M_G$.  After integrating out these
heavy neutrinos, we obtain the light neutrino mass matrix given by
\begin{equation} {\cal M} = U_e^{tr} \; [ \ m_\nu \ (V^{tr})^{-1} \ M_N  \ V^{-1} \ m_\nu^{tr} \ ] \;
U_e .
\end{equation}
It is explicitly defined in the lepton flavor basis where $U_e$ is
the $3\times3$ unitary matrix for left-handed leptons needed to
diagonalize $Y_e$ (eqn. \ref{eq:yukawa}), i.e. $Y_e^D = U_e^{tr} \
Y_e \ U_{\bar e}^*$.

 Note,
\begin{equation}(V^{tr})^{-1} = \frac{\hat M}{v_{16}} \left( \begin{array}{ccc}
 - 1/ (\tilde \phi^2 \ \xi) & 1/ \phi^1  & 0 \\
1/ \tilde \phi^2  & 0 & 0 \\
0 & 0 & 1/ \theta
\end{array} \right) .\end{equation}
Let us now define
\begin{equation} D^{tr} \equiv  m_\nu \ (V^{tr})^{-1} \ M_N \ {\cal P} = \left(
\begin{array}{cc}
a & 0  \\
a^\prime  & b  \\
0 & b^\prime
\end{array} \right)  \end{equation}
where \begin{equation} {\cal P} = \left(
\begin{array}{cc} 1 & 0
\\ 0 & 1 \\ 0 & 0 \end{array} \right) \end{equation} and $m_\nu$
is the Dirac neutrino mass (Eqn. \ref{eq:mnu}).  We then obtain
\begin{equation} {\cal M} =  U_e^{tr} \; [  \ D^{tr} \ \hat M_N^{-1} \ D \ ] \;
U_e \end{equation} with \begin{equation} \hat M_N \equiv \left(
\begin{array}{cc} M_1 & 0
\\ 0 & M_2 \end{array} \right) . \end{equation}  The
parameters $ \{ a, \ a^\prime, \ b, \ b^\prime \} $ are given by
\begin{eqnarray} a \equiv & - \epsilon^\prime \ \omega \ \lambda \ (M_1/
\tilde \phi^2) \ \frac{\hat M }{ v_{16}} \frac{v
\sin\beta}{\sqrt{2}} &  \\
a^\prime \equiv & ( -\epsilon^\prime \ \xi^{-1} + 3 \ \tilde
\epsilon ) \ \omega \ \lambda \ (M_1/ \tilde \phi^2) \ \frac{\hat
M }{v_{16}} \frac{v \sin\beta}{\sqrt{2}} & \nonumber \\
b \equiv & \epsilon^\prime \ \omega \ \lambda \ (M_2/ \phi^1) \
\frac{\hat M}{ v_{16}} \frac{v \sin\beta}{\sqrt{2}} & \nonumber \\
b^\prime \equiv & - 3 \ \epsilon \ \xi \ \sigma \ \lambda \ (M_2/
\phi^1) \ \frac{\hat M }{ v_{16}} \frac{v \sin\beta}{\sqrt{2}} . &
\nonumber
\end{eqnarray}  Note, these parameters are in general all complex.
We have  $ b \sim b^\prime$, since $\epsilon^\prime \sim \phi^1/
\hat M \sim \epsilon \ \xi \sim 3 \ \tilde \epsilon \ \xi$ (see
Eqns. \ref{eq:omega} and \ref{eq:brt}). However, without
fine-tuning, we find $ a \ll a^\prime$. This can be remedied if
the implicit order one coefficients in these expressions are used
to fine tune the two terms in $a^\prime$ so that they cancel to
about one part in 10 giving $ a^\prime \sim a$. Finally, due to
the small mixing angles in $U_e$, the observable neutrino mixing
matrix $U$ used to diagonalize ${\cal M}$ via
\begin{equation} {\cal M}^D = U^{tr} \ {\cal M} \ U \end{equation}
is given by \begin{equation} U = U_e^\dagger \ U_{FGY}
\end{equation} where $U_{FGY}$ is the FGY neutrino mixing matrix
(Eqn. \ref{eq:ufgy}). Using the results of Bla\v{z}ek et al.
(2000) (see Table 2 and discussion in Sect. 4.1)
\cite{Blazek:1999ue} we have
\begin{equation} |Y_e| \approx \left(
\begin{array}{ccc}
0 &  0.003 & 0.004 \\ 0.003 & 0.03 &  0.12 \\
0.004 & 0.12 & 1 \end{array} \right) \ 0.8  \label{eq:brt}
\end{equation} which gives $ |(U_e^{tr})_{1 2}| \approx 0.16$, $
|(U_e^{tr})_{2 3}| \approx 0.12$ and $ |(U_e^{tr})_{1 3}| \approx
0.01$.  Hence, the small mixing angles in $U_e$ do not
significantly affect the result of bi-large mixing.

In the $SO(10)$ version of the theory,  the discrete ambiguity
related to the mass ordering of the heavy Majorana neutrinos is
resolved with $M_1/M_2 \sim 10^3$ where we have used the relation
$m_{\nu_2}/m_{\nu_3} \approx (m_e/m_\mu) \ (M_1/M_2) \ \tilde
\epsilon $.  However since the heavy Majorana sector, including
the fields $ \{ \bar \nu_i; \; N_i, \; {\rm for} \; i=1,2,3 \} $
is much more complicated than studied previously
\cite{Raidal:2002xf}, and because of the additional CP violating
phases in this model, it is not clear what constraints may be
derived by requiring that the observed cosmological density come
via leptognesis.

However, the main point of FGY is that CP violation, as measured
in low energy neutrino oscillation experiments or via the
cosmological baryon asymmetry, is governed by the single phase
$\delta$ given in the expression  $ a^\prime \ e^{i \delta}$ (Eqn.
\ref{eq:phase}). In the minimal supersymmetric example we retain
this nice feature. In the $SO(10)$ example, we lose this simple
connection since there are generically additional CP violating
angles present in $U_e$. Note, however, that now there is the
possibility of relating CP violation in neutrino physics to CP
violation in the CKM matrix.

In this letter we have presented a ``natural" framework for the
FGY \cite{Frampton:2002qc} neutrino mass matrix ansatz.  The FGY
ansatz has two major virtues,
\begin{itemize}
\item it accomodates bi-large neutrino mixing, and \item it
relates CP violation in neutrino oscillations with the sign of the
cosmological baryon asymmetry.
\end{itemize}
We show that an $SU(2) \times U(1)$ family symmetry can provide
the desired FGY neutrino mass ansatz in a minimal supersymmetric
standard model.  Moreover the discrete ambiguity for the ratio of
Majorana neutrino masses is resolved.  This then makes definite
predictions for the sign of the CP violating phase $\delta$, and
also for lepton flavor violating processes (Eqn. \ref{eq:flavor}).
We have also shown how to obtain an approximate FGY ansatz in an
$SO(10)$ SUSY GUT. In this latter example we lose the direct
connection between high energy CP violation made visible by
leptogenesis and direct CP violation observable in low energy
neutrino oscillations. However since the neutrino sector is now
intertwined with the charged fermions, we have other interesting
relations. For example, the relation $b \approx b^\prime$, needed
for a large atmospheric neutrino mixing angle, is now automatic.
The large solar mixing angle is, however, still easily
accomodated. Moreover, the same $SU(2) \times U(1)$ family
symmetry has the virtue of generating the fermion mass hierarchy
as well as suppressing large flavor changing neutral current
processes \cite{Barbieri:1996ww,Blazek:1999ue}. Finally, it is
important to note that our results are not significantly affected
by renormalization group [RG] running.  It is well known that RG
running of the effective dimension 5 lepton-Higgs interaction
below the Majorana mass scales $[M_1, \ M_2] $ does not
significantly affect the results for neutrino masses and mixing
angles when we have a hierarchical neutrino
spectrum\cite{Chankowski:2001mx}.

 {\bf Acknowledgements}

Partial support for this work was obtained under DOE grant
DOE/ER/01545-840 and also a grant from the Alexander von Humboldt
Foundation. I would like to thank the faculty and staff at the
Universit\"{a}t Bonn where this work was begun.  I also greatly
benefitted from discussions with P.H. Frampton, J.E. Kim, H.D.
Kim, L. Schradin and K. Tobe.

%

\end{document}